# Equilibrium Dynamics of the Dissipative Two–State System

T. A. Costi[‡] and C. Kieffer
*Universität Karlsruhe, Institut für Theorie der Kondensierten Materie, 76128 Karlsruhe, Germany*

Wilson's momentum shell renormalization group method is used to solve for the dynamics of the dissipative two–state system. We utilize the mapping of the spin–boson model onto the anisotropic Kondo model (AKM) and solve for the dynamics of the latter. We find that the AKM captures the physics of the dissipative two–state system for dissipation strength $0 \leq \alpha \leq 4$ corresponding to $\infty \geq J_{\parallel} \geq -\infty$ in the AKM. The dynamics of the AKM shows a smooth crossover between two strong coupling regimes corresponding to weak and strong dissipation in the spin–boson model.

PACS numbers: 71.27.+a,75.20.Hr,71.10.+x,72.15.Qm

A problem which is of general interest in both physics and chemistry is that of a quantum mechanical system tunneling between two states and subject to a dissipative coupling to environmental degrees of freedom. Examples of this type of system abound; they include the tunneling of defects in metallic glasses [1], the motion of the total flux in a SQUID between two metastable fluxoid states [2] and the diffusion of protons and muons in metals [3]. The main theoretical interest lies in a description of the dynamics of the generalized coordinate of the two level system subject to the influence of the environment. Both the equilibrium and non–equilibrium dynamics are of interest for the different experimental realizations of two–level systems. In the case of macroscopic quantum coherence experiments in SQUIDs, the system can be prepared in one of the two states by applying a bias ( an external magnetic field) for times $t < 0$ and then allowed to evolve for $t > 0$ in zero bias. The non–equilibrium correlation functions of the two state system are then of primary interest. For most microscopic systems such an initial state preparation is not realizable and the interest then lies in the equilibrium dynamics. In this paper we present results for the equilibrium case with a bias.

Specifically we consider the spin–boson Hamiltonian,

$$H_{SB} = -\frac{1}{2}\hbar\Delta\sigma_x + \frac{1}{2}\epsilon\sigma_z + \sum_\alpha \omega_\alpha(a_\alpha^\dagger a_\alpha + \frac{1}{2}) + \frac{1}{2}q_0\sigma_z \sum_\alpha \frac{C_\alpha}{\sqrt{2m_\alpha\omega_\alpha}}(a_\alpha + a_\alpha^\dagger). \quad (1)$$

Here $\sigma_i, i = x, y, z$ are Pauli spin matrices, the two–states of the system correspond to $\sigma_z =\uparrow$ and $\sigma_z =\downarrow$. $\Delta$ is the bare tunneling matrix element and $\epsilon$ is a bias. The environment is represented by an infinite set of harmonic oscillators (labelled by the index $\alpha$) with masses $m_\alpha$ and frequency spectrum $\omega_\alpha$ coupling linearly to the coordinate $Q = \frac{1}{2}q_0\sigma_z$ of the two–level system via a term characterized by the couplings $C_\alpha$. The environment spectral function is given in terms of these couplings, oscillator masses and frequencies by $J(\omega) = \frac{\pi}{2}\sum_\alpha(\frac{C_\alpha^2}{m_\alpha\omega_\alpha})\delta(\omega - \omega_\alpha)$. In the case of an Ohmic heat bath, of interest to us here, we have $J(\omega) = 2\pi\alpha\omega$, for $\omega << \omega_c$, where $\omega_c$ is a high energy cut–off and $\alpha$ is a dimensionless parameter characterizing the strength of the dissipation.

A great deal of work has been carried out over the last 10 years in order to understand the dynamics of this apparently simple model [4,5]. Extensive calculations based on the Feynman–Vernon path integral formalism within the "Non–Interacting Blip Approximation" (NIBA) have yielded reliable information for weak dissipation and short times [4] and have also provided some insight into the expected behaviour in other regimes. In addition to such direct attempts at calculating dynamics for the spin–boson model, it has proved fruitful to exploit analogies between this model and several other models, including the inverse–square Ising model [6], the anisotropic Kondo model (AKM) and the Vigmann–Finkel'stein (VF) model [7]. In the long time approximation it has been shown that the partition functions, and hence the thermodynamics, of these different models can be put into correspondence and the parameters of the models related. This correspondence has also been carried out for the respective fermionic Hamiltonians by applying an approximate bosonization method valid for low energies $\omega << \omega_c$ [8]. In this paper we exploit such a mapping of the spin–boson model onto the AKM in order to make predictions about the dynamics of the former on the basis of renormalization group calculations on the latter. It has not been clear [4] to what extent the above mapping is valid in all parameter regimes, in particular for weak dissipation, and this has partly motivated this work. The AKM is given by

$$H = \sum_{k,\sigma}\epsilon_k c_{k\sigma}^\dagger c_{k\sigma} + \frac{J_\perp}{2}\sum_{kk'}(c_{k\uparrow}^\dagger c_{k'\downarrow}S^- + c_{k\downarrow}^\dagger c_{k'\uparrow}S^+) + \frac{J_\parallel}{2}\sum_{kk'}(c_{k\uparrow}^\dagger c_{k'\uparrow} - c_{k\downarrow}^\dagger c_{k'\downarrow})S^z + g\mu_B h S_z, \quad (2)$$

where the first term represents non–interacting conduction electrons and the second and third terms represent an exchange interaction between a localized spin 1/2 and the conduction electrons with strength $J_\perp, J_\parallel$. For $J_\perp = J_\parallel$ the model reduces to the usual Kondo model of magnetic impurities in metals. A local magnetic field, $h$, coupling only to the impurity spin in the Kondo model (the last term in Eq. 2) corresponds to a finite bias, $\epsilon$, in the spin–boson model. The correspondence between $H$ and $H_{SB}$ is then given by $\epsilon = g\mu_B h$, $\frac{\Delta}{\omega_c} = \rho J_\perp$ and $\alpha = (1+\frac{2\delta}{\pi})^2$, where $\tan\delta = -\frac{\pi\rho J_{||}}{4}$, $\delta$ is the phase shift for scattering of electrons from a potential $J_{||}/4$ and $\rho = 1/2D$ is the conduction electron density of states per spin at the Fermi level for a flat band of width $2D$ [4]. We note that whereas the spin–boson model can describe dissipation with arbitrary strength, $0 \leq \alpha \leq \infty$, the AKM is only capable of describing the region $0 \leq \alpha \leq 4$ with $\alpha = 0$ corresponding to $J_{||} = +\infty$, $\alpha = 1$ corresponding to the antiferromagnetic/ ferromagnetic boundary of the AKM ($J_{||} = 0$), and $1 < \alpha \leq 4$ corresponding to the ferromagnetic regime of the AKM.

The relevant dynamical quantity for the two–level system is the response function $\chi_{SB}(\omega, T) = \langle\langle\sigma_z;\sigma_z\rangle\rangle$ which translates into the local dynamic spin susceptibility, $\chi(\omega,T) = \langle\langle S_z; S_z\rangle\rangle$, for the Kondo model. In order to calculate this quantity we apply Wilson's momentum shell renormalization group method which has recently been generalized to the calculation of dynamical quantities for a number of models (e.g. [11]). Briefly, the procedure, explained in detail in [9], consists of (i) linearizing the spectrum about the Fermi energy $\epsilon_k \to v_F k$, (ii) introducing a logarithmic mesh of $k$ points $k_n = \Lambda^{-n}$ and (iii) performing a unitary transformation of the $c_{k\sigma}$ such that $f_{0\sigma} = \sum_k c_{k\sigma}$ is the first operator in a new basis, $f_{n\sigma}, n = 0, 1, \ldots$, which tridiagonalizes $H_c = \sum_{k\mu}\epsilon_{k\mu}c^\dagger_{k\mu}c_{k\mu}$ in k–space, i.e. $H_c \to \sum_\mu\sum_{n=0}^{\infty}\xi_n\Lambda^{-n/2}(f^\dagger_{n+1\mu}f_{n\mu} + h.c.)$, with $\xi_n \to (1+\Lambda^{-1})/2$ for $n >> 1$. The Hamiltonian (2) with the above discretized form of the kinetic energy is now diagonalized by the following iterative process: (a) one defines a sequence of finite size Hamiltonians $H_N = \sum_\mu\sum_{n=0}^{N-1}\xi_n\Lambda^{-n/2}(f^\dagger_{n+1\mu}f_{n\mu} + h.c.) + \frac{J_\perp}{2}(f^\dagger_{0\uparrow}f_{0\downarrow}S^- + f^\dagger_{0\downarrow}f_{0\uparrow}S^+) + \frac{J_{||}}{2}(f^\dagger_{0\uparrow}f_{0\uparrow} - f^\dagger_{0\downarrow}f_{0\downarrow})S^z$ for $N \geq 0$; (b) the Hamiltonians $H_N$ are rescaled by $\Lambda^{\frac{N-1}{2}}$ such that the energy spacing remains the same, i.e. $\bar{H}_N = \Lambda^{\frac{N-1}{2}}H_N$. This defines a renormalization group transformation $\bar{H}_{N+1} = \Lambda^{1/2}\bar{H}_N + \sum_\mu\xi_N(f^\dagger_{N+1\mu}f_{N\mu} + h.c.) - E_{G,N+1}$, with $E_{G,N+1}$ chosen so that the ground state energy of $\bar{H}_{N+1}$ is zero. Using this recurrence relation, the sequence of Hamiltonians $\bar{H}_N$ for $N = 0, 1, \ldots$ is iteratively diagonalized within a product basis of, typically, up to 1200 states for each iteration. This gives the excitations and many body eigenstates at a corresponding set of energy scales $\omega_N$ defined by $\omega_N = \Lambda^{-\frac{N-1}{2}}$ and allows a direct calculation of response functions. Thus, for example, $\chi(\omega,T)$ is given by $\chi(\omega,T) = \chi'(\omega,T) + i\chi''(\omega,T) = \frac{1}{Z_N}\sum_{m,n}|M^N_{m,n}|^2\frac{e^{-\beta\epsilon_m} - e^{-\beta\epsilon_n}}{\omega + i0 - (\epsilon_m - \epsilon_n)}$, where $\epsilon_m, \epsilon_n$ are many–body excitations of $H_N$, $Z_N(T)$ the corresponding partition function of $H_N$, and $M^N_{m,n} =< m|S_z|n >_N$ the relevant many–body matrix elements for the dynamic susceptibility. The quantity we actually calculate is $S(\omega) = -\frac{1}{\pi}\frac{\chi''(\omega+i\delta)}{\omega}$ which is related to the neutron scattering cross–section. Our results were obtained for $\Lambda = 2$, keeping the 320 lowest states at each iteration. In this paper we discuss only the $T = 0$ results.

The accuracy of the numerical calculations could be tested by (a) showing that the exactly solvable Toulouse limit, $\alpha = \frac{1}{2}$, could be reproduced (described below) and (b) by verifying the generalized Shiba relation for the dynamic spin susceptibility of the spin–boson model [12]. The latter relation states that, at $T = 0$, $S(\omega = 0) = 2\alpha\chi'(\omega = 0)^2$. Generalized with the extra factor of $\alpha$, it is also valid for the AKM, as explicitly verified by our numerical results. We chose to extract the static susceptibility via a Kramer's–Kronig relation $\chi'(\omega = 0) = \int_{-\infty}^{+\infty}S(\omega)d\omega$. Verified in this way, the generalized Shiba relation provides a good test of the method, not just at low frequency, but at all energy scales and for arbitrary values of $\alpha$ and $\Delta$. The error in $\chi'(0)$ was typically 5% and contributed the main source of error, approximately 10%, to the Shiba relation; representative cases are shown in Table I [13]. In all cases we found that $\chi''(\omega) \sim \omega$ for $\omega \to 0$ showing that the spin–spin correlation function, $\langle[\sigma_z(t), \sigma_z(0)]\rangle$, decays as $1/t^2$ for long times i.e. the tunneling is always incoherent at long times. In contrast, the NIBA gives an algebraic decay $\sim 1/t^{2(1-\alpha)}$ depending on the coupling $\alpha$.

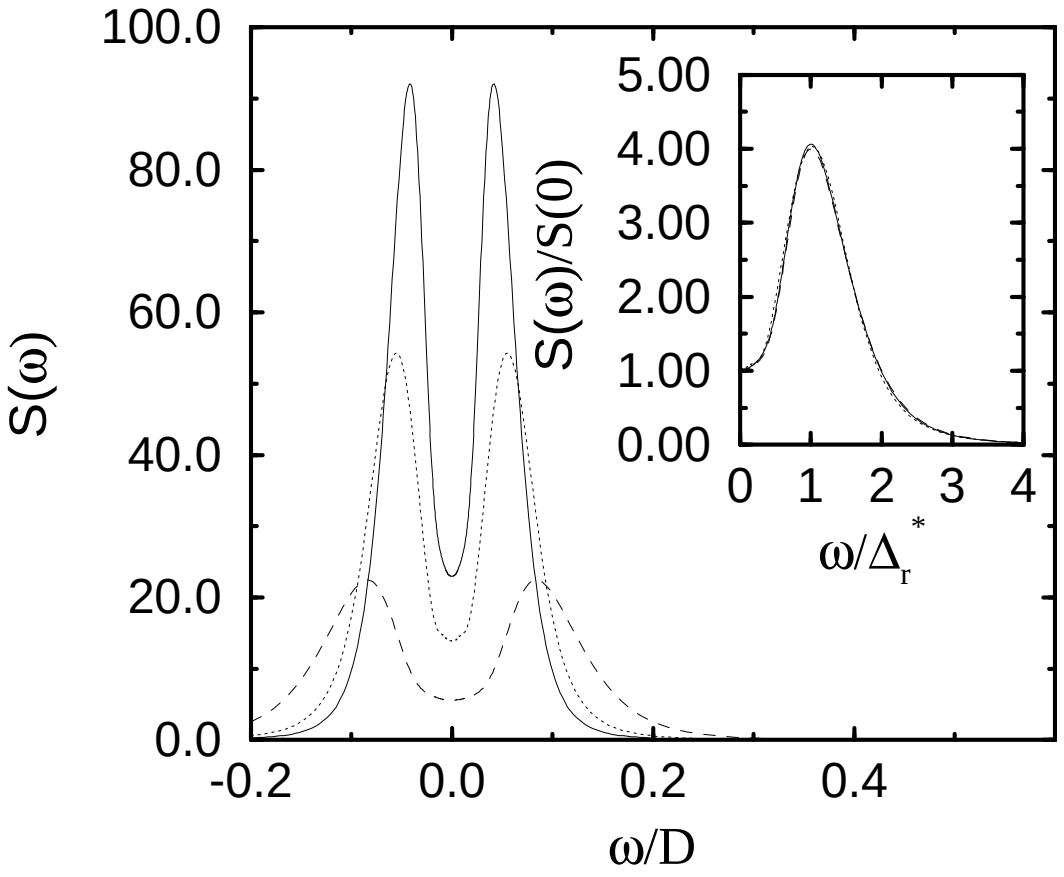

FIG. 1. The response function $S(\omega)$ of the AKM for $\alpha = 0.1$ ($J_{||} = 4.698$) and $\Delta = 0.08$ (solid line), $\Delta = 0.1$ (dotted line), $\Delta = 0.15$ (dashed line). The position of the inelastic peaks is at $\Delta_r^* = 0.045, 0.063$ and $0.087$ respectively. The inset shows that $S(\omega)/S(0)$ is a universal function of the rescaled energy $\omega/\Delta_r^*$ ($S(0) \sim \frac{1}{\Delta_r^{*2}}$). Energies are in units of $D = \frac{\omega_c}{2} = 1$.

The case of zero dissipation, $\alpha = 0$, corresponds to $J_{||} = \infty$ in the AKM. This is exactly solvable; $S(\omega) = \frac{1}{4\omega}[\delta(\omega - J_\perp) - \delta(\omega + J_\perp)]$ is a sum of two delta functions at $\omega = \pm J_\perp$. In the spin–boson model the response is also a sum of two delta functions at $\omega = \pm\Delta$. We can thus identify the cut–off $\omega_c$ appearing in $\frac{\Delta}{\omega_c} = \rho J_\perp$ as the bandwidth, $\omega_c = 2D$. In this case coherent oscillations with frequency $\Delta$ are realized. On increasing $\alpha$, Fig. (1), the above peaks survive but are broadened. The system exhibits oscillations for short times $t \lesssim 1/\Delta_r$, where $\Delta_r$ is the renormalized tunneling frequency (see below). As discussed above, the long time behaviour is always incoherent for any finite $\alpha$ and $\Delta$. The position of the peaks, which is a measure of the renormalized tunneling frequency, is reduced i.e. as expected, dissipation hinders tunneling. Fig. (1) illustrates these features for $\alpha = 0.1$. For $\alpha << 1$ and small $\Delta$, scaling arguments give a renormalized tunneling frequency $\Delta_r = \Delta(\frac{\Delta}{\omega_c})^{\frac{\alpha}{1-\alpha}}$ [4]. Numerical results for $\alpha = 0.01$ gave for the inelastic peak positions, $\Delta_r^*$, values within 5% of this result for $\Delta = 0.05, 0.07$ and 0.1 [13]. Although, the amplitude of the damped oscillations decreases with increasing $\Delta$, the rescaled spin response $S_\alpha(\omega)/S_\alpha(0)$ is a universal function of $\omega/\Delta_r^*$ (depending on $\alpha$). This is shown in the inset to Fig. (1) for $\alpha = 0.1$. Hence, for weak dissipation, increasing $\Delta$, for $\Delta < \omega_c$, does not destroy the damped oscillations. As we show below, only increasing $\alpha$ has this effect.

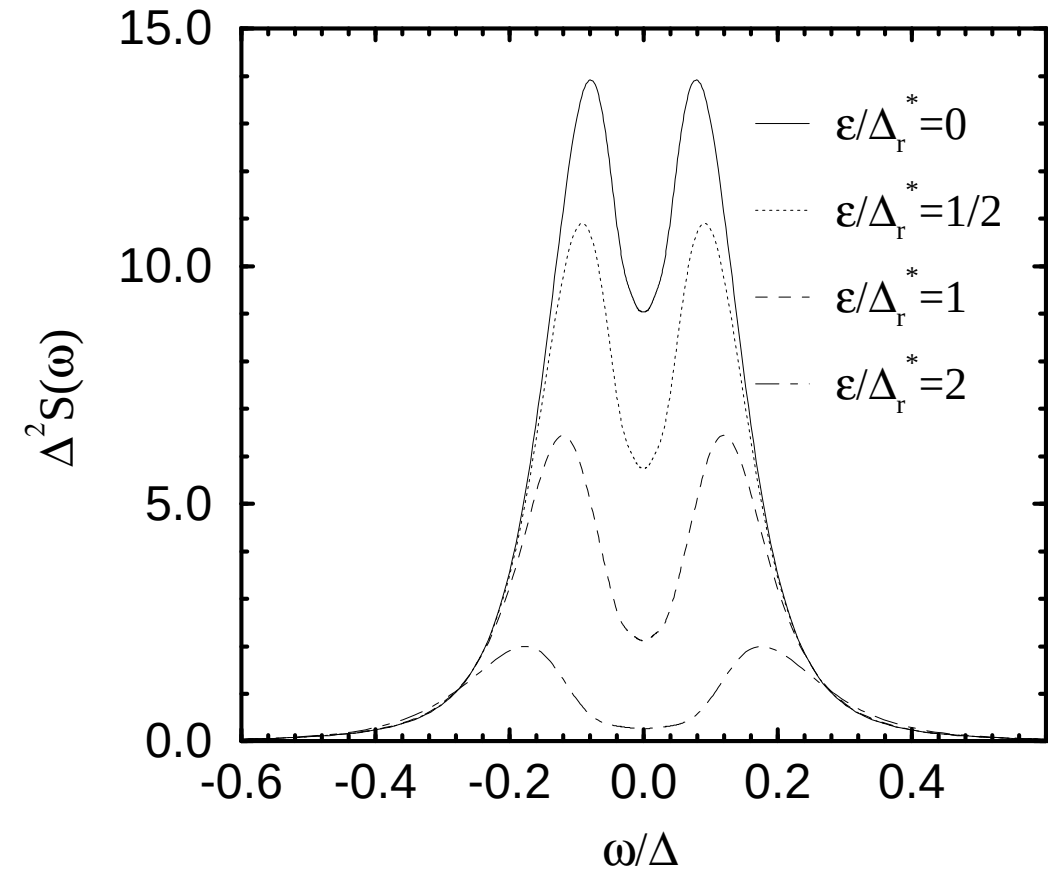

FIG. 2. $S(\omega)$ for $\alpha = 0.2$ ($J_\| = 3.01$), $\Delta = 0.001$ and different bias $\epsilon/\Delta_r^*$. $\Delta_r^* = 0.079\Delta$ is the renormalized tunneling amplitude for the zero bias case. For the finite bias cases, $\epsilon/\Delta_r^* = \frac{1}{2}, 1, 2$, we find for the renormalized tunneling amplitude $\Delta_b^*/\Delta_r^* = 1.05, 1.33, 1.93$, respectively, which is within 6% agreement of the weak coupling result [5] $\Delta_b = \sqrt{\Delta_r{}^{*2} + \epsilon^2}$. Energies are in units of $D = \frac{\omega_c}{2} = 1$.

Fig. (1) for $\alpha = 0.1$ and the zero bias case of Fig. (2) for $\alpha = 0.2$ show that the damped oscillations become strongly suppressed with increasing $\alpha$, even for $\Delta << 1$.

The oscillations disappear completely at approximately $\alpha = 0.33$ [14], well below the Toulouse point $\alpha = \frac{1}{2}$ (see below). A small bias, $\epsilon << \Delta$, i.e. small relative to the bare tunneling amplitude, but comparable to the renormalized tunneling amplitude, $\Delta_r^*$, leads to a suppression of the inelastic peaks, and an increase in the tunneling frequency relative to the zero bias value, Fig. (2). This increase in the effective tunneling frequency, $\Delta_b(\epsilon)$, is well described by the weak coupling theory expression [5], $\Delta_b(\epsilon) = \sqrt{\epsilon^2 + \Delta_r^{*2}}$. The effect of a bias, however, is not simply to renormalize the tunneling frequency, since we could not scale $S(\omega)$ for different $\epsilon$ onto a single curve, as is the case for different tunneling amplitudes and zero bias. Thus a finite bias for *weak* dissipation does not change the qualitative aspects of the dynamics. Beyond $\alpha \approx 0.2$ the inelastic peaks are negligible, even for very small values of $\Delta$, and the incoherent part dominates (see also Fig. (4)).

For the exactly solvable Toulouse limit, $\alpha = \frac{1}{2}$, the numerical results for $S(\omega)$ fit very well onto the resonant level model result for all values of $\Delta$ as shown in Fig. (3).

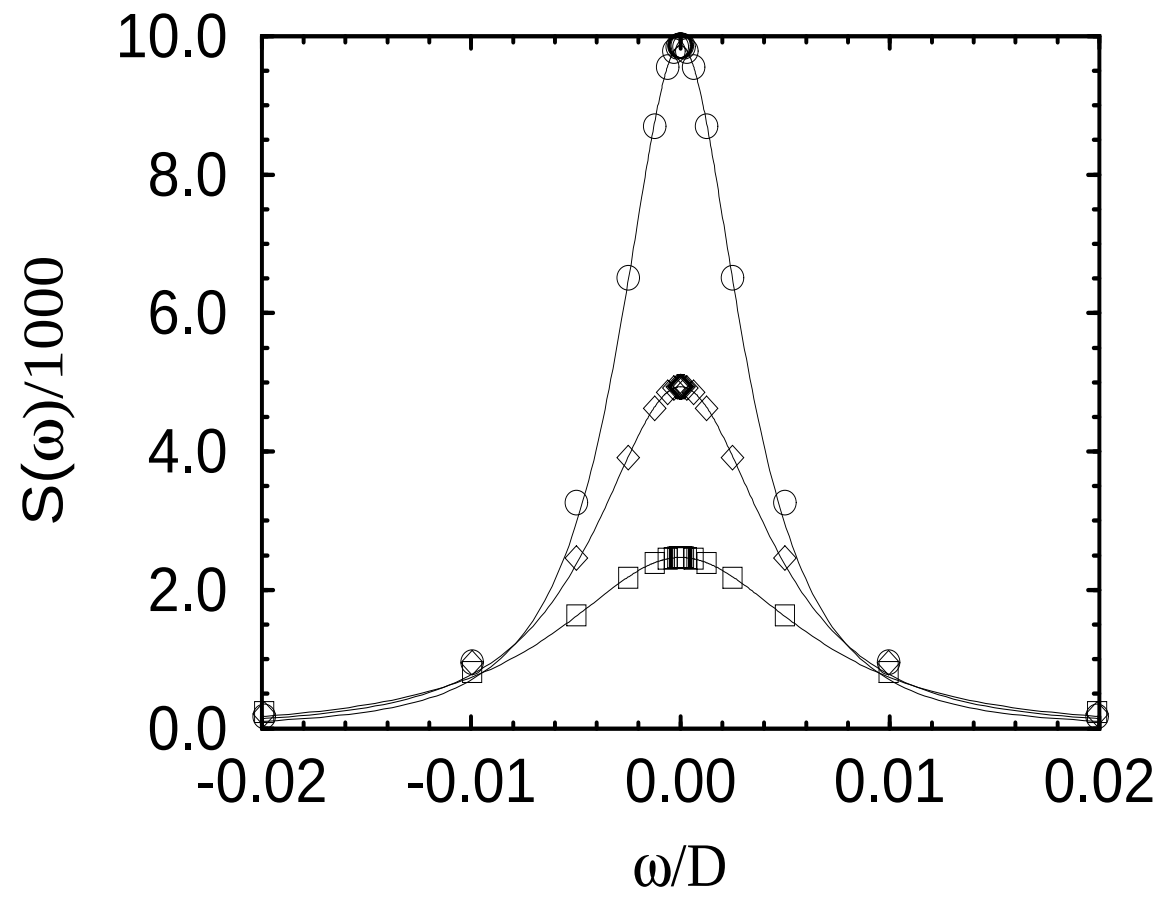

FIG. 3. $S(\omega)$ for the exactly solvable Toulouse case, $\tilde{\alpha} = \frac{1}{2}$, ($\tilde{J}_\| = 1.312$). $\Delta = 0.1$ (circles), $\Delta = 2^{\frac{1}{4}}0.1$ (diamonds), $\Delta = 2^{\frac{1}{2}}0.1$ (squares). The solid lines are the exact analytic results for the resonant level model.

The response consists of a single peak at $\omega = 0$ of width $\tilde{\Delta}_{RL} = \pi\rho V_{RL}^2$ where $V_{RL}\omega_{RL}^{-1} = \frac{J_\perp}{2}\omega_{KM}^{-1}\cos^2(\delta)$ and $\omega_{RL}, \omega_{KM}$ are the respective high–energy cut–offs of the resonant level and Kondo models [7]. Since $\omega_{KM} = 2D = \omega_c$, we can relate the various models by determining $\omega_{RL}$ from the exact resonant level model result $S(\omega = 0) = \frac{1}{\pi^2\tilde{\Delta}^2}$. We found that $\frac{\omega_{RL}}{\omega_{KM}} = 1.125$ in all cases with a variation of less than 0.1% [13].

The regime $\frac{1}{2} < \alpha < 1$ corresponds to the antiferromagnetic Kondo regime, for which there have been few reliable results for the dynamic susceptibility. The renormalization group results presented here provide an essentially exact solution in this region, as can be seen from Table I. The single quasielastic peak in $S(\omega)$ narrows exponentially for $\alpha \to 1$, i.e. $J_{\perp,\|} \to 0$, (Fig. (4)). Tunneling is incoherent for all values of $\Delta$. Fig. (4) shows the crossover from damped oscillations for weak dissipation to incoherent relaxation for strong dissipation. The crossover occurs at approximately $\alpha = 0.33$. From the scaling properties of $S_\alpha(\omega)$ for different $\Delta$ this value is independent of $\Delta$ (for $\Delta << \omega_c$). The crossover point separates two strong–coupling regimes, corresponding to

(a) damped short time oscillations for small $\alpha$, and (b) incoherent relaxation of the spin and Kondo behaviour for large $\alpha$. The details of the fixed points and the behaviour in the ferromagnetic regime $J_{\|} < 0$, $-J_{\|} > J_{\perp}$, ($\alpha > 1$), where we obtain the expected localization of the particle, will be discussed elsewhere. The inset to Fig. (4) shows that the low energy scale is given correctly by the scaling result, $\Delta_r$, for weak dissipation and by the Bethe ansatz expression for the Kondo temperature of the AKM [10] for $\alpha \to 1$.

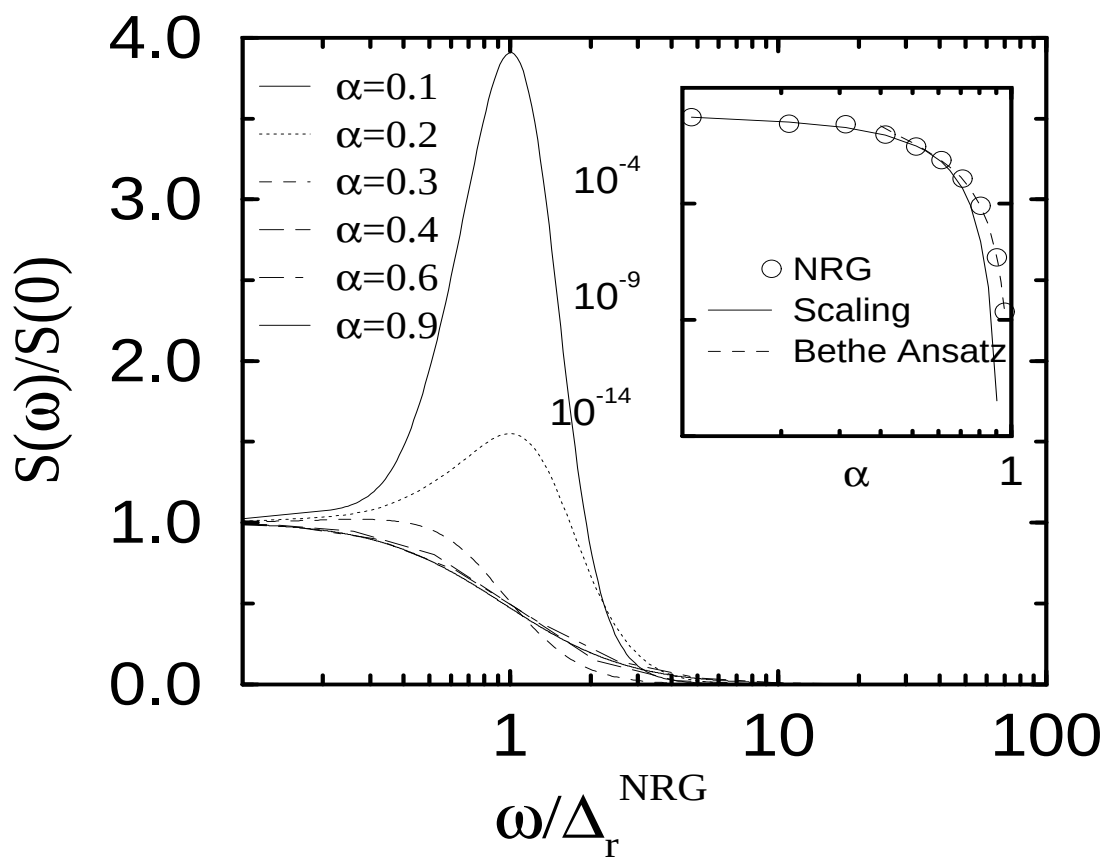

FIG. 4. $S(\omega)$ for $\Delta = 0.1$ and different values of $\alpha$. The inelastic peak disappears at approximately $\alpha = 0.33$. The low energy scale, $\Delta_r^{NRG}$, for $\Delta = 0.1$ and for $\alpha = 0.1, 0.2, \ldots, 0.9$ is shown in the inset. It is defined to be the position of the inelastic peak, $\Delta_r^*$, for small $\alpha$ and the half–width of the quasielastic peak for large $\alpha$. It agrees with the scaling result, $\Delta_r$, for small $\alpha$, and with the Bethe ansatz Kondo temperature for the AKM [10], $T_K(J_\perp, J_\|)$, for $\alpha \to 1$. At $\alpha \approx 0.3$, close to the crossover, there is an ambiguity in defining the low energy scale and some discrepancy is seen. Similarly, $T_K(J_\|, J_\perp)$ deviates from the true energy scale, $\Delta_r^{NRG}$, when $\alpha < 0.5$ ( $T_K(J_\|, J_\perp)$ in [10] is derived for small $J_\|$ i.e. $\alpha \to 1$).

To summarize, we have shown, that at low energies, $0 \leq \omega \lesssim \Delta << \omega_c$, the equilibrium dynamics of the dissipative two–state system is very well described by the AKM for dissipation $0 \leq \alpha \leq 1$. Agreement with exact results was obtained in the limiting cases $\alpha \to 0, \frac{1}{2}, 1$. The validity of the generalized Shiba relation, within errors consistent with the numerical data, was verified in the range $0 < \alpha < 1$ and $\Delta << \omega_c$.

We acknowledge useful discussions with A. Rosch, P. Wölfle and T. Kopp. This work was supported by E.U. grants CT92-0068 (TAC), CT93-0115 (TAC, CK).

‡ Present address: Institut Laue-Langevin, B.P.156 38042 Grenoble, Cedex 9, France.

TABLE I. The Shiba relation, $S(0) = 2\tilde{\alpha}\chi'(0)^2$, for the dynamic spin susceptibility; $\tilde{\alpha}$ is defined in [13].

| $\alpha$ | $\Delta = J_\perp$ | $J_\|$ | $\tilde{\alpha}$ | $2\tilde{\alpha}[\chi'(0)]^2$ | $[-\frac{\chi''(\omega)}{\pi\omega}]_{\omega\to 0}$ | %error |
|---|---|---|---|---|---|---|
| 0.01 | 0.001 | 16.078 | 0.0108 | $6.536 \times 10^3$ | $6.450 \times 10^3$ | 1.3% |
| 0.1 | 0.01 | 4.698 | 0.1068 | $2.181 \times 10^3$ | $2.420 \times 10^3$ | 10.9% |
| 0.2 | 0.1 | 3.008 | 0.2111 | $7.095 \times 10^2$ | $7.638 \times 10^1$ | 7.4% |
| 0.4 | 0.001 | 1.659 | 0.4144 | $1.372 \times 10^{10}$ | $1.291 \times 10^{10}$ | 6.4% |
| 0.5 | 0.025 | 1.262 | 0.5139 | $3.748 \times 10^6$ | $3.965 \times 10^6$ | 5.5% |
| 0.8 | 0.1 | 0.4262 | 0.8071 | $8.568 \times 10^8$ | $7.919 \times 10^8$ | 8.2% |
| 0.9 | 0.1 | 0.2057 | 0.9037 | $2.262 \times 10^{13}$ | $2.437 \times 10^{13}$ | 7.7% |